%% file: main.tex
\documentclass{IEEEtran}
\usepackage{booktabs}
\usepackage[table]{xcolor}
\definecolor{best}{RGB}{220,50,50}
\definecolor{second}{RGB}{255,165,80}
\definecolor{third}{RGB}{255,230,100}
\input{macros}

\newcommand{\equalcontrib}{\textsuperscript{\ensuremath{\mathsection}}}

\title{A Sheaf-Theoretic Framework for Distributed Multi-Site Channel Charting%
\thanks{This work was supported by the SNS JU project 6G-GOALS under the EU Horizon Europe
program, Grant Agreement No. 101139232, and by Huawei Technology France SASU under
Grant No. Tg20250616041.}%
}

\author{
\IEEEauthorblockN{
Enrico Grimaldi\equalcontrib\IEEEauthorrefmark{1}\IEEEauthorrefmark{2},~
Leonardo Di Nino\equalcontrib\IEEEauthorrefmark{2}\IEEEauthorrefmark{3},~
Mario Edoardo Pandolfo\equalcontrib\IEEEauthorrefmark{1}\IEEEauthorrefmark{2},\\~
Gabriele D'Acunto\IEEEauthorrefmark{2}\IEEEauthorrefmark{3},~
Sergio Barbarossa\IEEEauthorrefmark{3}~and~
Paolo Di Lorenzo\IEEEauthorrefmark{2}\IEEEauthorrefmark{3}
}
\\
\IEEEauthorblockA{%
\IEEEauthorrefmark{1}Department of Computer, Control and Management Engineering,
Sapienza University of Rome, Rome, Italy}\\
\IEEEauthorblockA{%
\IEEEauthorrefmark{2}National Inter-University Consortium for Telecommunications (CNIT),
Parma, Italy}\\
\IEEEauthorblockA{%
\IEEEauthorrefmark{3}Department of Information Engineering, Electronics, and Telecommunications,
Sapienza University, Rome, Italy}
\\
\IEEEauthorblockA{%
E-mail:\{name.surname\}@uniroma1.it}. \equalcontrib These authors contributed equally to this work. \vspace{-.4cm}}

\begin{document}
\maketitle
\begin{abstract}
Channel charting (CC) enables data-driven user localization in wireless networks by embedding channel state information (CSI) into low-dimensional representations. In multi-cell scenarios, each base station independently learns a local chart via neural encoders, leading to misaligned representation spaces across overlapping coverage areas. This lack of consistency hinders network-level tasks such as user tracking, handover prediction, and resource allocation. To address this issue, we propose a principled framework for multi-site channel charting based on topological signal processing. We model the collection of local charts as a network sheaf, which encodes consistency constraints across the network and enables the coherent integration of locally learned representations into a shared global structure. This formulation introduces an interpretable inductive bias that promotes alignment across base stations while preserving local geometric fidelity. Building on this model, we develop a multi-site channel charting architecture and an alternating optimization algorithm that jointly updates neural encoders and inter-site orthogonal transport maps, with theoretical guarantees on consistency. Experimental results validate the effectiveness of the proposed approach, demonstrating improved cross-site alignment without degrading the quality of local embeddings. 
\end{abstract}
\begin{keywords}
Channel charting, topological signal processing,
sheaves, distributed representation learning.
\vspace{-.2cm}
\end{keywords}
\input{sections/01_introduction}

\input{sections/02_model}

\input{sections/03_results}
\input{sections/04_conclusions}
\begingroup
\balance
\bibliographystyle{bibstile}
\bibliography{bibliography}
\endgroup
\end{document}

%% file: macros.tex
\usepackage[english]{babel}

\usepackage{cite}
\usepackage{type1cm} 
\usepackage{graphicx} 
\usepackage{xspace} 
\usepackage{balance} 
\usepackage{booktabs} 
\usepackage{multirow} 
\usepackage[font={bf}, tableposition=top]{caption} 
\usepackage{subcaption} 
\usepackage{bold-extra} 
\usepackage{algorithm}
\usepackage{algorithmic}
\usepackage{microtype} 
\usepackage{siunitx} 
\usepackage{xfrac} 
\usepackage{amsmath}
\usepackage{amssymb}
\usepackage{mathtools}
\usepackage{amsfonts}
\usepackage{amsthm}
\usepackage{bbm}
\usepackage[hyphens]{url} 
\usepackage[bookmarks, pdftex, colorlinks=true]{hyperref} 
\usepackage{csquotes}
\usepackage[show]{chato-notes} 
\usepackage{bm}
\usepackage{xspace}
\usepackage{physics}
\usepackage{newunicodechar}
\newunicodechar{⟓}{\ensuremath{\uplus}}
\usepackage{tikz}
\usetikzlibrary{positioning}
\usetikzlibrary{bayesnet}
\usetikzlibrary{arrows}
\usetikzlibrary{shapes}
\usetikzlibrary{fit}
\usepackage{tikz-cd}
\usepackage{bbold}
\usetikzlibrary{calc, positioning, arrows.meta, shapes.geometric, fit, backgrounds}
\usepackage{pgf}
\usepackage{tikz-3dplot}
\usepackage{tcolorbox}

\newcommand{\spara}[1]{\smallskip\noindent\textbf{#1}}

\usepackage[dvipsnames]{xcolor}
\hypersetup{
   bookmarks, pdftex,
   colorlinks=true,
   pagebackref=true, backref=page,
   linkcolor={red!50!black},
   filecolor={green!50!black},
   citecolor={green!50!black}, 
   urlcolor={blue!80!black},
}

\graphicspath{{fig}}

\theoremstyle{plain}

\newtheorem{definition}{Definition}

\usepackage{stackengine}

\definecolor{mypurple}{RGB}{254, 68, 218}

%% file: sections/01_introduction.tex
\section{Introduction}
Channel Charting (CC) \cite{ferrand2023wireless} has emerged as a powerful framework for addressing the high dimensionality of Channel State Information (CSI) in MIMO radio access systems, while preserving its intrinsic geometric structure. By leveraging dimensionality reduction techniques ranging from linear methods to neural approaches \cite{studer2018channel}, CC maps high-dimensional CSI measurements into compact latent representations that capture the underlying spatial relationships of the propagation environment.
Extending CC to multi-cell scenarios, however, introduces new challenges. 
Charts learned independently from local channel observations may lack global consistency, particularly in regions where coverage overlaps. 
Ensuring aligned representations across cells, such as consistency at anchor points and continuity of user trajectories across cell boundaries, is essential for enabling reliable downstream tasks.
In this work, we establish a representation learning framework for network-wide channel charting that enforces global coherence. 
To this end, we leverage tools from cellular sheaf theory \cite{curry2014sheaves,hansen2020laplacians}, which provide a principled way to encode and enforce consistency constraints across distributed network entities.\\
\textbf{Related works.} Multi-point channel charting has been studied in \cite{deng2018multipoint} as a data fusion problem, where features collected by multiple base stations observing the same environment are coherently combined. Extensions of this setting have also been explored in federated frameworks \cite{agostini2022federated}.
In contrast, multi-site channel charting addresses the more challenging scenario in which sensing regions only partially overlap, requiring the alignment of representation spaces across different base stations.
Prior work \cite{vindas2024multi} approached this problem through optimal transport theory, whereas we instead formulate it using sheaf-theoretic tools \cite{curry2014sheaves,hansen2020laplacians}.
Sheaf-based methods have recently gained attention as a principled framework for modeling and optimizing distributed systems \cite{hansen2019distributed,hanks2025distributed}, with applications ranging from federated learning \cite{issaid2025tackling} to network formation in semantic communications \cite{grimaldi2025learning}.
Our approach differs from these works in a fundamental way. 
In \cite{issaid2025tackling}, sheaves are used as structural priors, acting as inductive biases in the parameter space. In contrast, we operate directly at the level of representation spaces. 
While \cite{grimaldi2025learning} employs sheaf-based formulations for network-level alignment, it relies on pre-trained neural agents; in contrast, we jointly train the neural encoders together with the synchronization mechanism.\\
\textbf{Contributions.} In this paper, we develop a principled framework for multi-site channel charting that enforces global consistency across distributed representations. To this end, we make the following contributions. First, we introduce a network sheaf formulation for multi-site channel charting and connect it to the spectral theory of sheaves, yielding an inductive bias that ensures consistent global representations. Second, building on this formulation, we cast the joint learning problem as a regularized optimization that couples neural encoders with inter-site transport maps, and we propose an efficient distributed alternating optimization scheme, where base stations update their local models through message passing and perform closed-form updates for the alignment step. Third, we validate the proposed framework on standard channel charting benchmarks, demonstrating improved cross-site alignment while preserving local geometric fidelity and outperforming existing baselines.

%% file: sections/02_model.tex
\section{Sheaf-theoretic Channel Charting}
\spara{Background.} Network sheaves provide a mathematical framework to equip the nodes of a graph with local data structures, while enforcing global consistency through constraints on edges \cite{curry2014sheaves,hansen2020laplacians}. In this work, we focus on network sheaves with vector space assignments.
\begin{definition}[Network Sheaf]
Given a graph $\mathcal{G}(\mathcal{V},\mathcal{E})$, a network sheaf $\mathcal{F}(\mathcal{G})$ consists of:
\begin{itemize}
    \item a vector space $\mathcal{F}(v)$ for each node $v \in \mathcal{V}$ (node stalk);
    \item a vector space $\mathcal{F}(e)$ for each edge $e \in \mathcal{E}$ (edge stalk);
    \item linear maps $\rho_u^{\mathcal{F}(e)}:\mathcal{F}(u)\rightarrow\mathcal{F}(e)$ and $\rho_v^{\mathcal{F}(e)}:\mathcal{F}(v)\rightarrow\mathcal{F}(e)$ for each edge $e\sim(u,v)$ (restriction maps).
\end{itemize}
\end{definition}
This structure encodes a local-to-global consistency principle: data are assigned at nodes and constrained to agree across edges. Global consistency is captured by the notion of global section.

\begin{definition}[Global Sections]
Given a network sheaf $\mathcal{F}(\mathcal{G})$, the space of \textbf{global sections} $\mathcal{H}^0_{\mathcal{F}} \subseteq \bigoplus_{v\in\mathcal{V}}\mathcal{F}(v)$ consists of assignments of local data to nodes simultaneously satisfying all edge-wise consistency constraints:
\begin{equation}
    \mathbf{s}\in\mathcal{H}^0_{\mathcal{F}} \;\Leftrightarrow\; 
    \rho_u^{\mathcal{F}(e)}\mathbf{s}_u = \rho_v^{\mathcal{F}(e)}\mathbf{s}_v,\quad \forall\, e = (u,v) \in \mathcal{E}.
\end{equation}
\end{definition}

\begin{figure}[t]
    \centering
    \includegraphics[width=1\linewidth]{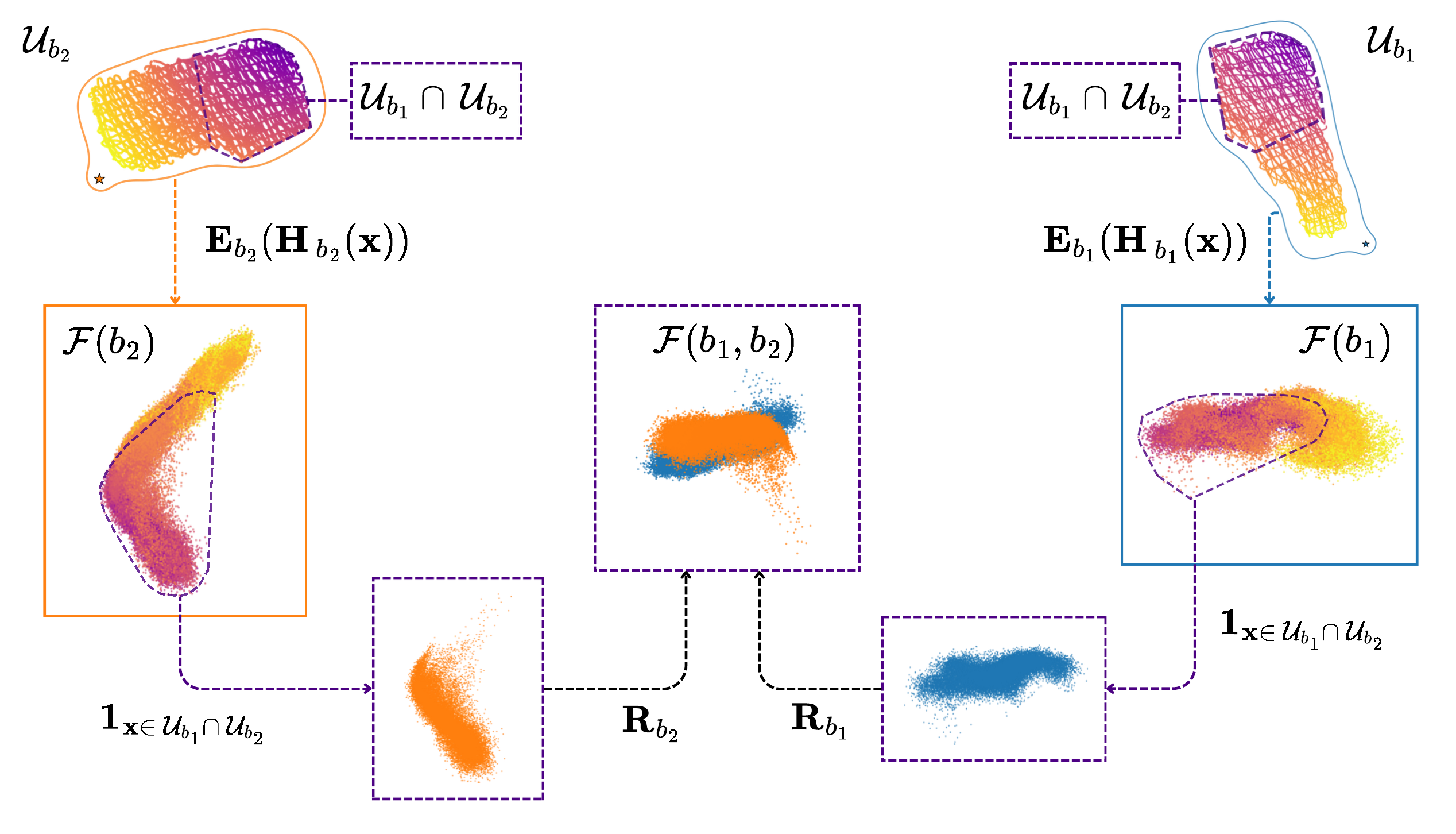}
    \caption{Visualization of an edge of the proposed network sheaf as a composition of a domain restriction and an isometry locally defined on the nodes of the network.}
    \label{fig:sheafCC}
\end{figure}

\spara{System model.} We consider a sensing environment $\boldsymbol{\Omega}$ and a certain number $B$ of base stations $\mathcal{B}=\{b_1,...,b_B\}$ such that their coverage areas $\{\mathcal{U}_b\}_{b \in \mathcal{B}}$ satisfy $\boldsymbol{\Omega}=\bigcup_{b\in\mathcal{B}} \mathcal{U}_{b}$.
We further assume that the coverage regions are not disjoint, i.e., that there exists at least one pair of base stations $(b_i,b_j)$ such that $\mathcal{U}_{b_i}\cap\,\mathcal{U}_{b_j}\neq \emptyset$.
A single-antenna user moves within $\boldsymbol{\Omega}$ and is observed at discrete time instants. 
At each time step, the user transmits a signal over $F$ subcarriers, and the base stations estimate CSI.
Each base station is equipped with a neural encoder $\mathbf{E}_b:\mathbb{C}^{R_b\times F}\rightarrow\mathbb{R}^n$ that maps CSI to channel chart embeddings, where $R_b$ denotes the number of receiver antennas at the base station and $n=2$ is typically chosen to ensure spatial interpretability. The guiding principle of a single-site channel charting pipeline is that CSIs measured from points being near in space should be mapped into embeddings being near in the representation space associated with the encoder performing dimensionality reduction. We extend this idea to multi-site channel charting, introducing a \textit{gluing} principle such that CSIs measured from points being in overlapping coverages should be mapped to the same points in latent spaces of the encoders associated to the base stations of interest. 
To this aim, we introduce a \textit{multi-site channel charting network sheaf} specified as follows:
\begin{itemize}
    \item For each base station $b \in \mathcal{B}$, we associate a node with a corresponding node stalk:
    \begin{equation}
        \mathcal{F}(b)=\{\mathbf{E}_b(\mathbf{H}_b(\mathbf{x}))\in\mathbb{R}^n:\mathbf{x}\in\mathcal{U}_b\}
    \end{equation}
    collecting the embeddings of the CSI corresponding to points within the coverage area of base station $b$;
    \item For each pair $(b_i,b_j): \mathcal{U}_{b_i}\cap\mathcal{U}_{b_j}\neq \emptyset$, we associate an edge with a corresponding edge stalk:
   \begin{equation}
    \mathcal{F}(b_i,b_j)=\{\mathbf{E}_*(\mathbf{H}_*(\mathbf{x}))\in\mathbb{R}^n:\mathbf{x} \in \mathcal{U}_{b_i}\cap\mathcal{U}_{b_j}\}
   \end{equation}
    collecting the CSI embeddings for points in the overlap of the coverage areas;
    \item On each edge, we define the following restriction maps:
     \begin{align}
\rho^{\mathcal{F}(b_i,b_j)}_{b_i}\mathbf{E}_{b_i}(\mathbf{H}_{b_i}(\mathbf{x})) 
&= \mathbf{R}_{b_i}^{(b_i,b_j)}\mathbf{E}_{b_i}(\mathbf{H}_{b_i}(\mathbf{x})) \, \mathbf{1}_{\mathbf{x}\in\mathcal{U}_{b_i}\cap\mathcal{U}_{b_j}} \nonumber\\
\rho^{\mathcal{F}(b_i,b_j)}_{b_j}\mathbf{E}_{b_j}(\mathbf{H}_{b_j}(\mathbf{x})) 
&= \mathbf{R}_{b_j}^{(b_i,b_j)}\mathbf{E}_{b_j}(\mathbf{H}_{b_j}(\mathbf{x})) \, \mathbf{1}_{\mathbf{x}\in\mathcal{U}_{b_i}\cap\mathcal{U}_{b_j}} \nonumber
\end{align}
where $\mathbf{R}_{b_i}^{(b_i,b_j)}, \mathbf{R}_{b_j}^{(b_i,b_j)} \in \mathrm{O}(n)$ are orthogonal transformations acting as isometries on the learned embeddings over the overlapping region.
\end{itemize}
The proposed model can be interpreted as the composition of a discrete covering sheaf \cite{robinson2017sheaves}, capturing the restriction of data to overlapping coverage regions, and an $\mathrm{O}(n)$ bundle, modeling the orthogonal transformations that align local embeddings across base stations, as illustrated in Fig.~\ref{fig:sheafCC}.

Orthogonal restriction maps \cite{bodnar2022neural,barbero2022sheaf} transport the
embeddings across the edges of the sheaf without distorting their geometry,
allowing the signal to propagate back and forth without information loss.
The gluing property requires that local embeddings agree across overlapping coverage regions. This is captured by the global sections of the proposed sheaf, i.e., ${\{\mathbf{E}_b(\mathbf{H}_b(\mathbf{x}))\}_{b\in\mathcal{B}} \in \mathcal{H}^0_{\mathcal{F}}}$, satisfying $\forall (b_i,b_j),\ \mathbf{x}\in\mathcal{U}_{b_i}\cap\mathcal{U}_{b_j}$:
\begin{equation}
\rho^{\mathcal{F}(b_i,b_j)}_{b_i}\mathbf{E}_{b_i}(\mathbf{H}_{b_i}(\mathbf{x})) =
\rho^{\mathcal{F}(b_i,b_j)}_{b_j}\mathbf{E}_{b_j}(\mathbf{H}_{b_j}(\mathbf{x})),
\end{equation}

Ideally, we would like to leverage this system model as an inductive bias to steer the learning trajectory of the encoders at each base station to satisfy this gluing principle. However, as defined, the sheaf does not guarantee the existence of non-trivial global sections. This issue can be addressed by leveraging the spectral characterization of sheaves. In particular, non-trivial global sections are ensured if the edge connections, defined by the isometries $\mathbf{R}_{b_i}^{(b_i,b_j)}, \mathbf{R}_{b_j}^{(b_i,b_j)}$, admit the following factorization:
\begin{subequations}\label{eq:connection_factorized}
\begin{gather}
\mathbf{R}_{b_i}^{(b_i,b_j)} = \mathbf{R}_{b_j}^\top \mathbf{R}_{b_i} \\ 
\mathbf{R}_{b_j}^{(b_i,b_j)} = \mathbf{I}
\end{gather}
\end{subequations}
where $\mathbf{R}_{b_i}, \mathbf{R}_{b_j} \in \mathrm{SO}(n)$ denote local reference frames associated with each base station \cite{chung2014ranking}. 
This shift from $\mathrm{O}(n)$ bundles to \textit{flat bundles} \cite{di2026learning,bambergerbundle} not only guarantees the existence of global sections, but also reduces the complexity of the problem. Indeed, optimizing edge-wise maps over a graph incurs a worst-case space complexity of $\mathcal{O}(B^2 n^2)$, whereas the proposed reparameterization scales linearly with the number of base stations, i.e., $\mathcal{O}(B n^2)$.\\
\textbf{Problem formulation.} We parametrize each encoder by $\theta_b$ and define the index set $\mathcal{I}=\{(\mathbf{x},b_i,b_j):\mathbf{x}\in\mathcal{U}_{b_i}\cap\mathcal{U}_{b_j}\}$. The multi-site channel charting problem with the sheaf inductive bias is formulated as
\begin{align}\label{eq:global_problem}
    &\underset{\{\theta_b,\mathbf{R}_b\in \mathrm{SO}(n)\}_{b\in\mathcal{B}}}{\mathrm{min}} \quad 
    \sum_{b\in\mathcal{B}}\mathcal{L}_b(\theta_b) \nonumber\\  
    &\;\;+\lambda \hspace{-.3cm}\sum_{(\mathbf{x},b_i,b_j)\in\mathcal{I}}
    \norm{\mathbf{R}_{b_i}\mathbf{E}_{b_i}(\mathbf{H}_{b_i}(\mathbf{x}))-\mathbf{R}_{b_j}\mathbf{E}_{b_j}(\mathbf{H}_{b_j}(\mathbf{x}))}_\mathrm{F}^2 .
\end{align}
The term $\mathcal{L}_b(\theta_b)$ is an unsupervised representation learning loss.
In particular, we adopt the triplet loss formulation\cite{ferrand2021triplet}, which enforces local geometric consistency by bringing nearby points closer while separating distant ones. 
Given an anchor sample $\mathbf{x}_a$ acquired at time $t_a$, we define
positive samples $\mathbf{x}_p$ as temporally coherent with the anchor,
i.e., $|t_p - t_a| \le \Delta T$ with $\Delta T = 5\,\mathrm{s}$, and negative
samples $\mathbf{x}_n$ as those outside this window, $|t_n - t_a| > \Delta T$.
For a fixed margin $m>0$, the resulting loss is
\begin{align}
    \mathcal{L}(\theta_b)=\sum_{(\mathbf{x}_a,\mathbf{x}_p,\mathbf{x}_n)}\Big[&\|\mathbf{E}_{b}(\mathbf{H}_{b}(\mathbf{x}_a))-\mathbf{E}_{b}(\mathbf{H}_{b}(\mathbf{x}_p))\|_\mathrm{F}^2\nonumber \\
    & \hspace{-1cm}-\|\mathbf{E}_{b}(\mathbf{H}_{b}(\mathbf{x}_a))-\mathbf{E}_{b}(\mathbf{H}_{b}(\mathbf{x}_n))\|_\mathrm{F}^2+m\Big]^+
\end{align}
The second term in \eqref{eq:global_problem} enforces the sheaf-theoretic gluing constraint by promoting agreement between embeddings over overlapping coverage regions. By exploiting the factorization in \eqref{eq:connection_factorized}, it decouples the alignment problem over the edges into node-wise reference frames synchronization.

\section{Distributed Algorithmic Solution}
To solve \eqref{eq:global_problem} in a distributed manner, we adopt a message passing protocol among base stations. At each iteration, every base station exchanges information with its neighbors $\tilde{b}\in\mathcal{N}(b)$. Specifically, base station $b$ receives from each neighbor $\tilde{b}$ the embeddings over their overlapping region, expressed in the neighbor's reference frame.
As a result, at each iteration, base station $b$ has access to the set
$\{\mathbf{R}_{\tilde{b}}\mathbf{E}_{\tilde{b}}(\mathbf{H}_{\tilde{b}}(\mathbf{x}))\,\mathbf{1}_{\mathbf{x}\in\mathcal{U}_b\cap\mathcal{U}_{\tilde{b}}}\}_{\tilde{b}\in\mathcal{N}(b)},$
which provides all the information required to perform the alternating optimization locally.

\spara{Local reference frame update.} 
At each base station $b$, the update of the local reference frame is given by:
\begin{equation}\label{eq:local_reference_problem}
    \underset{\mathbf{R}_b\in\mathrm{SO}(n)}{\mathrm{min}} \;
    \sum_{\tilde{b}\in\mathcal{N}(b)} 
    \sum_{\mathbf{x}\in\mathcal{U}_b\cap\mathcal{U}_{\tilde{b}}} 
    \norm{\mathbf{R}_{b}\mathbf{E}_{b}(\mathbf{H}_{b}(\mathbf{x}))-\mathbf{R}_{\tilde{b}}\mathbf{E}_{\tilde{b}}(\mathbf{H}_{\tilde{b}}(\mathbf{x}))}_F^2.
\end{equation}
We can easily derive a closed form solution for the problem \eqref{eq:local_reference_problem} by noticing that, given the orthogonality constraints on $\mathbf{R}_b$, the problem can be rewritten as the following generalized eigenvalue problem:
\begin{equation}\label{eq:local_reference_problem_reload}
\underset{\mathbf{R}_b\in\mathrm{SO}(n)}{\mathrm{max}} \mathrm{Tr}\Big\{\mathbf{R}_b\sum_{(b,\tilde{b})}\sum_{\mathbf{x}\in\mathcal{U}_b\cap\mathcal{U}_{\tilde{b}}}\mathbf{E}_{b}(\mathbf{H}_{b}(\mathbf{x}))[\mathbf{E}_{\tilde{b}}(\mathbf{H}_{\tilde{b}}(\mathbf{x}))]^\top\mathbf{R}_{\tilde{b}}^\top\Big\}
\end{equation}
for which the solution is given by the Kabsch algorithm \cite{kabsch1976solution}. Specifically, at iteration $k$, the local reference frame $\mathbf{R}_b$ is updated by solving \eqref{eq:local_reference_problem} via a Procrustes alignment. To this end, we consider the singular value decomposition of the cross-covariance matrix aggregated over the neighborhood:
\begin{align}
\mathbf{C}_b^{(k)} &=
\sum_{\tilde{b}\in\mathcal{N}(b)} 
\sum_{\mathbf{x}\in\mathcal{U}_b\cap\mathcal{U}_{\tilde{b}}}
\mathbf{E}_{b}(\mathbf{H}_{b}(\mathbf{x}))
\left[\mathbf{E}_{\tilde{b}}(\mathbf{H}_{\tilde{b}}(\mathbf{x}))\right]^\top
(\mathbf{R}_{\tilde{b}}^{(k)})^\top \nonumber\\
&= \mathbf{U}\mathbf{\Sigma}\mathbf{V}^\top.
\end{align}
The update is then given by:
\begin{equation}
\mathbf{R}_{b}^{(k+1)} = \mathbf{U}\,\tilde{\mathbf{\Sigma}}\,\mathbf{V}^\top\in\mathrm{SO}(n),
\end{equation}
where $\tilde{\mathbf{\Sigma}}=\mathrm{diag}(1,\dots,1,\mathrm{det}(\mathbf{U}\mathbf{V}^\top)).$\\
\textbf{Encoder parameters update.}
Given the reference frames $\{\mathbf{R}_b^{(k+1)}\}_{b\in\mathcal{B}}$, the encoder parameters $\{\theta_b\}_{b\in\mathcal{B}}$ are updated by minimizing \eqref{eq:global_problem} with respect to $\theta_b$. This results in a gradient-based update, where each base station performs a local optimization step using its own data and the embeddings received from neighboring nodes. In practice, the update is implemented via stochastic gradient descent (or its variants) and can be efficiently computed using automatic differentiation within standard deep learning frameworks. The resulting procedure alternates between closed-form updates of the reference frames and gradient-based updates of the encoder parameters.\\
\textbf{Remark.} The work in \cite{vindas2024multi} proposes a multi-site CC
framework based on manifold alignment via optimal transport
\cite{alvarez2019towards}. In contrast, our approach leverages tools from
algebraic topology together with a Procrustes-type formulation
\cite{wang2008manifold}. While the two approaches are related and thus
naturally comparable, ours provides stronger theoretical guarantees and a
more interpretable, robust inductive bias. Moreover, \cite{vindas2024multi}
relies on a fully neural parametrization over the base station graph edges,
whereas our formulation admits a closed-form alignment step that scales
linearly with the number of nodes. The resulting over-parameterization can
hinder training in optimal transport-based methods, particularly at large
scale, while our approach remains computationally efficient and easier to
optimize in a distributed fashion.

%% file: sections/03_results.tex
\section{Numerical Results}

\begin{figure}
    \centering
    \includegraphics[width=0.7\linewidth]{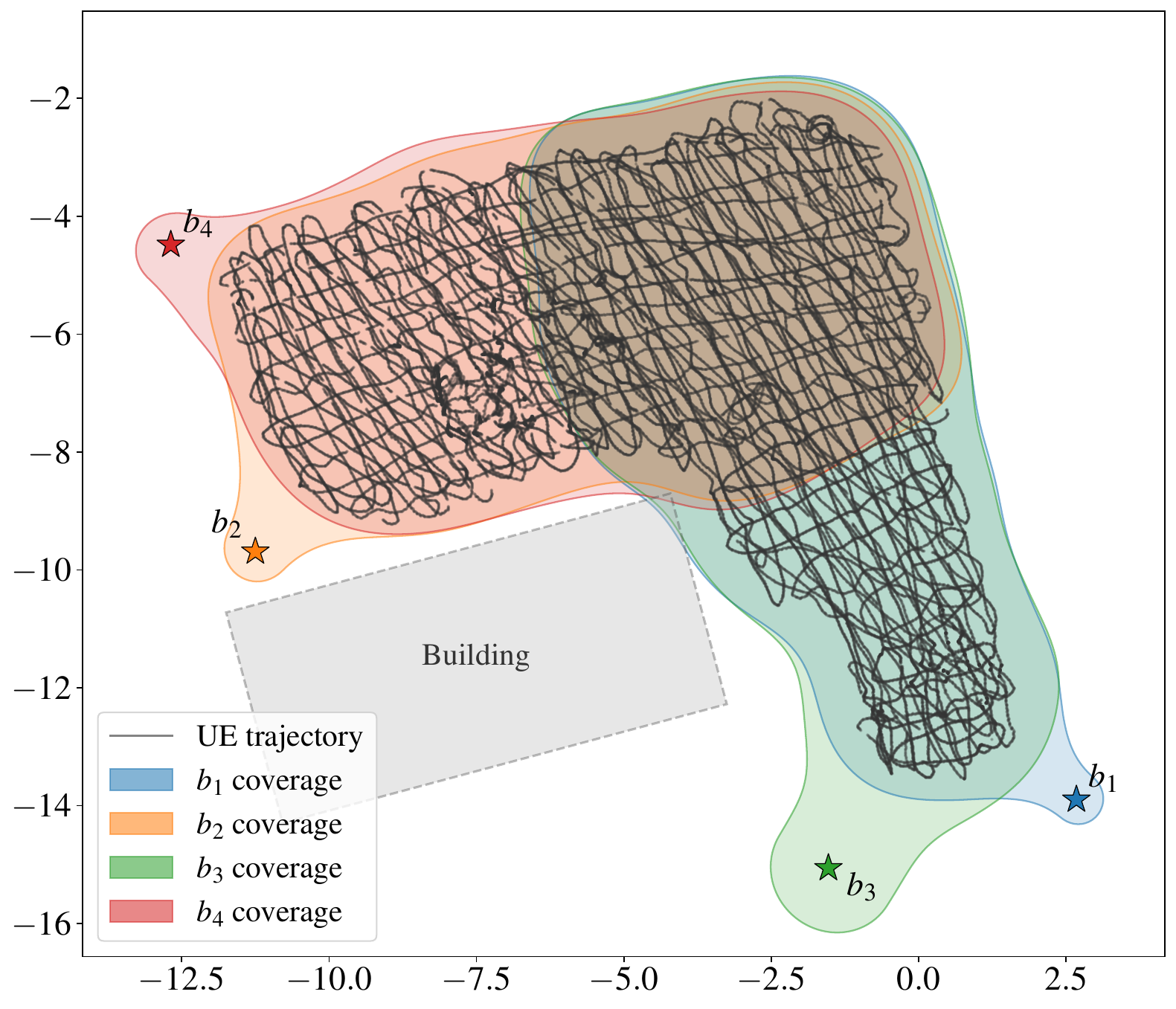}
    \caption{Visualization of the dataset and the partition of the sensing environment.}
    \label{fig:dichasus_map}
\end{figure}

\begin{figure}
    \centering
    \includegraphics[width=1\linewidth]{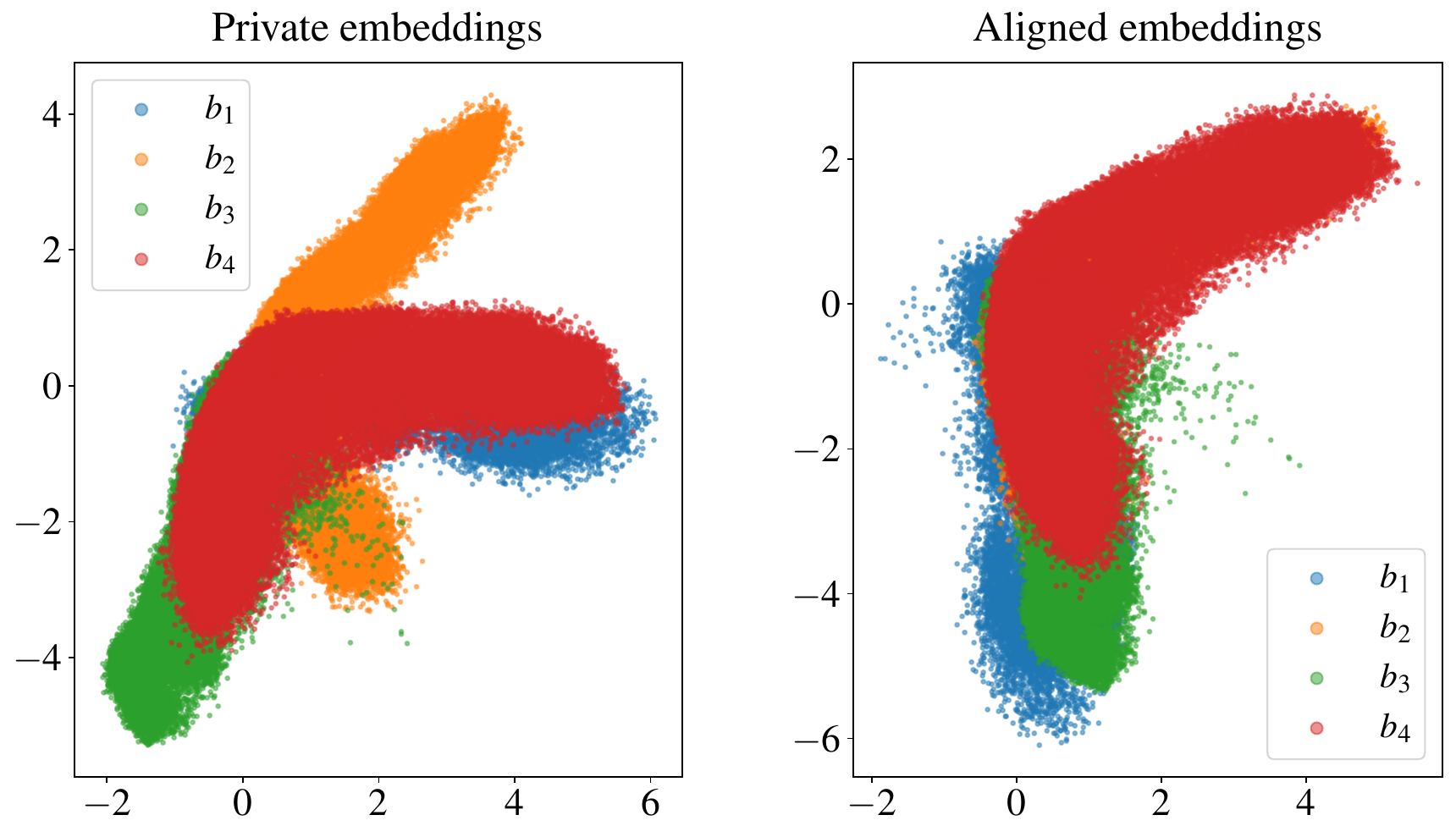}
    \caption{Learned embedding with flat bundle alignment and induced shared representation space.}
    \label{fig:aligned}
\end{figure}

\noindent\textbf{Dataset and system configuration.}
We evaluate the proposed framework on the DICHASUS dataset \cite{euchner2021distributed}, using the concatenation of the trajectories \texttt{dichasus\_cf02}, \texttt{dichasus\_cf03}, and \texttt{dichasus\_cf04}\footnote{The source code for the numerical evaluation is available at: \url{https://github.com/SPAICOM/sheaf-theoretic-channel-charting.git}}.
The four available access points are treated as distinct base stations, each observing the same propagation environment from a different spatial perspective.
Each base station is equipped with a neural encoder, preceded by a non-trainable preprocessing layer as in \cite{ferrand2021triplet}, to extract channel features from raw CSI measurements. The embedding dimension is fixed to $n=2$ to enable geometric interpretability of the learned charts. Coverage areas are defined as shown in Fig.~\ref{fig:dichasus_map} by partitioning the sensing region according to the presence of dominant scatterers, which results in partially overlapping coverage regions and thus provides the necessary conditions to study multi-site alignment.

\noindent\textbf{Baselines.}
We compare the proposed method against a set of representative baselines covering different alignment strategies:
\emph{(i)} \textit{unsupervised optimal transport} \cite{vindas2024multi}, which performs nonlinear alignment of latent spaces via neural mappings defined over graph edges;
\emph{(ii)} \textit{cover sheaf}, corresponding to the flat bundle with identity reference frames, where embeddings are directly constrained to coincide in overlapping regions;
\emph{(iii)} \textit{$\mathrm{O}(n)$ bundle}, i.e., the sheaf formulation without connection factorization, where edge-wise orthogonal maps are learned independently via Procrustes alignment \cite{di2024learning};
\emph{(iv)} \textit{vanilla}, where base station encoders are trained independently without any form of alignment; and
\emph{(v)} \textit{federated} methods, in both standard and personalized variants, where models are trained via parameter sharing across nodes.

\noindent\textbf{Metrics.}
We assess both local embedding quality and cross-site consistency. For local geometric fidelity, we report Kruskal Stress (KS) \cite{kruskal1964multidimensional}, Trustworthiness (TW), and Continuity (CT) \cite{venna2001neighborhood}, which quantify how well the learned embeddings preserve neighborhood relationships; all metrics are averaged across base stations. For global alignment, we adopt the FOSCTTM metric \cite{chen2023unsupervised}, which measures how well corresponding samples from different sites are matched in the latent space.

\noindent\textbf{Results.}
For reproducibility, all sheaf-theoretic algorithms impose a cosine schedule on the sheaf penalization term $\lambda$ in \eqref{eq:global_problem}, which is gradually increased to a predefined maximum value over the training epochs. All results are reported for a total of $20$ training epochs, using a decaying learning rate ($10^{-3}$ for the first $10$ epochs and $10^{-4}$ thereafter) for the stochastic gradient descent updates that minimize \eqref{eq:global_problem} during the encoder parameter update step.

The quantitative results in Table~\ref{tab:results} show that the proposed sheaf-based methods achieve competitive dimensionality reduction performance across all metrics. In particular, they match or slightly outperform the vanilla single-site baseline, while consistently improving over the optimal transport approach. Approaches relying on parameter averaging typically fail in multi-site channel charting: the federated approach encourages encoders to behave similarly from a functional perspective, whereas true comparability lies in the latent geometry of the sensing space, which should in turn shape the representation space. This suggests that injecting inductive bias directly at the latent-space level is a more effective strategy than enforcing consistency at the parameter level. From the perspective of local geometry preservation, the Bundle variant achieves the lowest KS and the highest TW, indicating the best reconstruction of the intrinsic structure of the sensing environment. The Flat Bundle attains nearly identical performance, confirming that the imposed alignment does not degrade local embedding quality. In contrast, federated baselines exhibit significantly higher KS and lower TW, highlighting the detrimental effect of parameter averaging on representation quality in this setting.

Regarding cross-site alignment, the Flat Bundle achieves the best FOSCTTM score, outperforming all competing methods. This result is particularly significant: despite relying only on orthogonal transformations with a closed-form Procrustes solution, it surpasses the optimal transport approach, which employs more expressive nonlinear mappings over graph edges. This behavior can be explained by the structure of the flat bundle: the factorization in \eqref{eq:connection_factorized} guarantees the existence of global sections, thereby inducing a consistent shared latent space across all base stations. In contrast, the $\mathrm{O}(n)$ bundle, which learns edge-specific transformations without such factorization, achieves inferior alignment despite its higher flexibility, suggesting that the additional degrees of freedom hinder optimization. The Cover Sheaf achieves intermediate alignment performance, confirming that enforcing agreement without learnable reference frames is too restrictive. The Vanilla baseline, while competitive in terms of KS and TW, fails to align representations across sites, emphasizing the need for explicit coupling mechanisms.

Finally, the Flat Bundle provides a globally consistent geometric representation, as all embeddings are expressed in a common reference frame $\{\mathbf{R}_b\}_{b\in\mathcal{B}}$, as shown in Fig.~\ref{fig:aligned}. In contrast, methods based on edge-wise mappings only ensure pairwise consistency and do not yield a unified global geometry. Overall, the Flat Bundle offers the best trade-off between alignment accuracy, complexity, and interpretability.

\begin{table}
  \centering
  \footnotesize
  \setlength{\tabcolsep}{3.5pt}
  \begin{tabular}{lrrccr}
    \toprule
    \multirow{2}{*}{\textbf{Method}} & \multirow{2}{*}{KS $\downarrow$} & \multirow{2}{*}{TW $\uparrow$}
      & \multicolumn{2}{c}{CT $\uparrow$} & \multirow{2}{*}{FOSCTTM $\downarrow$} \\
    \cmidrule(lr){4-5}
    & & & $K{=}2$ & $K{=}38$ & \\
    \midrule
    Bundle              & $\mathbf{0.386}$    & $\mathbf{0.834}$    & $\underline{0.964}$ & $\mathbf{0.927}$    & $0.295$ \\
    Cover Sheaf         & $0.408$             & $0.811$             & $0.950$             & $0.897$             & $\mathit{0.227}$ \\
    Flat Bundle         & $\underline{0.387}$ & $\underline{0.830}$ & $\mathit{0.961}$    & $\underline{0.925}$ & $\mathbf{0.154}$ \\
    Opt. Transport\cite{vindas2024multi} & $0.404$ & $\mathit{0.825}$ & $0.954$ & $0.915$ & $\underline{0.183}$ \\
    Federated           & $0.540$             & $0.695$             & $0.901$             & $0.830$             & $0.558$ \\
    Pers. Fed.\cite{agostini2022federated} & $\mathit{0.390}$ & $0.796$ & $0.946$ & $0.902$ & $0.500$ \\
    Vanilla             & $0.391$             & $0.823$             & $\mathbf{0.968}$    & $\mathit{0.924}$    & $0.449$ \\
    \bottomrule
  \end{tabular}
  \caption{Mean KS, TW, CT, and FOSCTTM. TW is reported for $K=10$ (approximately constant for $K\in[5,40]$); CT (continuity) is shown at local ($K{=}2$) and global ($K{=}38$) scales. $\mathbf{Bold}$, $\underline{underline}$, and $\mathit{italic}$ denote best, second, and third per column.}
  \label{tab:results}
\end{table}

%% file: sections/04_conclusions.tex

\section{Conclusions}
We proposed a sheaf-theoretic framework for multi-site channel charting that
provides a principled way to enforce consistency across distributed
representations. By casting the problem in terms of network sheaves, we
introduce a structured, interpretable inductive bias that links local
embeddings to a globally coherent latent space, with guarantees on the
existence of consistent global representations. The resulting formulation
leads to a distributed optimization scheme with closed-form alignment
updates and linear complexity in the number of base stations, making it
suitable for large-scale wireless networks. Numerical results show that the
proposed approach improves cross-site alignment while preserving local
geometric fidelity, consistently outperforming existing baselines; in
particular, the flat bundle construction emerges as a key design choice,
offering a favorable trade-off between alignment accuracy, computational
efficiency, and interpretability. Future work will extend the sheaf-based
modeling to more expressive encoder architectures and more complex
propagation environments, and integrate task-driven objectives for adaptive
network coordination and application-aware representation learning.

%% file: bibliography.bib
@inproceedings{agostini2022federated,
 author = {P. Agostini and Z. Utkovski and S. Sta{\'n}czak},
 booktitle = {IEEE 23rd International Workshop on Signal Processing Advances in Wireless Communication (SPAWC)},
 organization = {IEEE},
 pages = {1--5},
 title = {Federated learning for multipoint channel charting},
 year = {2022}
}

@inproceedings{alvarez2019towards,
 author = {D. Alvarez-Melis and S. Jegelka and T. S. Jaakkola},
 booktitle = {The 22nd International Conference on Artificial Intelligence and Statistics},
 organization = {PMLR},
 pages = {1870--1879},
 title = {Towards optimal transport with global invariances},
 year = {2019}
}

@inproceedings{bambergerbundle,
 author = {J. Bamberger and F. Barbero and X. Dong and M. M. Bronstein},
 booktitle = {The Thirteenth International Conference on Learning Representations},
 title = {Bundle Neural Network for message diffusion on graphs}
}

@inproceedings{barbero2022sheaf,
 author = {F. Barbero and C. Bodnar and H. S. de Oc{\'a}riz Borde and M. Bronstein and P. Veli{\v{c}}kovi{\'c} and P. Li{\`o}},
 booktitle = {Topological, Algebraic and Geometric Learning Workshops},
 organization = {PMLR},
 pages = {28--36},
 title = {Sheaf neural networks with connection laplacians},
 year = {2022}
}

@article{bodnar2022neural,
 author = {C. Bodnar and F. Di Giovanni and B. Chamberlain and P. Lio and M. Bronstein},
 journal = {Advances in Neural Information Processing Systems},
 pages = {18527--18541},
 title = {Neural sheaf diffusion: A topological perspective on heterophily and oversmoothing in gnns},
 volume = {35},
 year = {2022}
}

@inproceedings{chen2023unsupervised,
 author = {D. Chen and B. Fan and C. Oliver and K. Borgwardt},
 booktitle = {Eleventh International Conference on Learning Representations (ICLR )},
 title = {Unsupervised Manifold Alignment with Joint Multidimensional Scaling},
 year = {2023}
}

@article{chung2014ranking,
 author = {F. Chung and W. Zhao and M. Kempton},
 journal = {Internet Mathematics},
 pages = {87--115},
 publisher = {Taylor \& Francis},
 title = {Ranking and sparsifying a connection graph},
 volume = {10},
 year = {2014}
}

@book{curry2014sheaves,
 author = {J. M. Curry},
 publisher = {University of Pennsylvania},
 title = {Sheaves, cosheaves and applications},
 year = {2014}
}

@inproceedings{deng2018multipoint,
 author = {J. Deng and S. Medjkouh and N. Malm and O. Tirkkonen and C. Studer},
 booktitle = {Asilomar Conference on Signals, Systems, and Computers},
 pages = {286--290},
 title = {Multipoint channel charting for wireless networks},
 year = {2018}
}

@inproceedings{di2024learning,
 author = {L. Di Nino and S. Barbarossa and P. Di Lorenzo},
 booktitle = {58th Asilomar Conference on Signals, Systems, and Computers},
 organization = {IEEE},
 pages = {59--63},
 title = {Learning Sheaf Laplacian Optimizing Restriction Maps},
 year = {2024}
}

@inproceedings{di2026learning,
  title={Learning the structure of connection graphs},
  author={L. Di Nino and G. D’Acunto and S. Barbarossa and P. Di Lorenzo},
  booktitle={ICASSP 2026-2026 IEEE International Conference on Acoustics, Speech and Signal Processing (ICASSP)},
  pages={76--80},
  year={2026},
  organization={IEEE}
}

@inproceedings{euchner2021distributed,
 author = {F. Euchner and M. Gauger and S. D{\"o}rner and S. ten Brink},
 booktitle = {WSA ; 25th International ITG Workshop on Smart Antennas},
 organization = {VDE},
 pages = {1--6},
 title = {A distributed massive MIMO channel sounder for “big CSI data”-driven machine learning},
 year = {2021}
}

@article{ferrand2021triplet,
 author = {P. Ferrand and A. Decurninge and L. G. Ordonez and M. Guillaud},
 journal = {IEEE Journal on Selected Areas in Communications},
 number = {8},
 pages = {2361--2373},
 publisher = {IEEE},
 title = {Triplet-based wireless channel charting: Architecture and experiments},
 volume = {39},
 year = {2021}
}

@article{ferrand2023wireless,
 author = {P. Ferrand and M. Guillaud and C. Studer and O. Tirkkonen},
 journal = {IEEE Communications Magazine},
 number = {6},
 pages = {124--130},
 publisher = {IEEE},
 title = {Wireless channel charting: Theory, practice, and applications},
 volume = {61},
 year = {2023}
}

@inproceedings{grimaldi2025learning,
  title={Learning network sheaves for ai-native semantic communication},
  author={E. Grimaldi and M.E. Pandolfo and G. D’Acunto, Gabriele and S. Barbarossa and P. Di Lorenzo},
  booktitle={2025 59th Asilomar Conference on Signals, Systems, and Computers},
  pages={1692--1696},
  year={2025},
  organization={IEEE}
}

@inproceedings{hanks2025distributed,
 author = {T. Hanks and H. Riess and S. Cohen and T. Gross and M. Hale and J. Fairbanks},
 booktitle = {IEEE Conference on Decision and Control},
 pages = {3057--3064},
 title = {Distributed multi-agent coordination over cellular sheaves},
 year = {2025}
}

@inproceedings{hansen2019distributed,
 author = {J. Hansen and R. Ghrist},
 booktitle = {57th annual allerton conference on communication, control, and computing (allerton)},
 organization = {IEEE},
 pages = {565--571},
 title = {Distributed optimization with sheaf homological constraints},
 year = {2019}
}

@phdthesis{hansen2020laplacians,
 author = {J. Hansen},
 school = {University of Pennsylvania},
 title = {Laplacians of cellular sheaves: Theory and applications},
 year = {2020}
}

@article{issaid2025tackling,
 author = {C. B. Issaid and P. Vepakomma and M. Bennis},
 journal = {arXiv preprint arXiv:2502.01145},
 title = {Tackling feature and sample heterogeneity in decentralized multi-task learning: A sheaf-theoretic approach},
 year = {2025}
}

@article{kabsch1976solution,
 author = {W. Kabsch},
 journal = {Foundations of Crystallography},
 number = {5},
 pages = {922--923},
 publisher = {International Union of Crystallography},
 title = {A solution for the best rotation to relate two sets of vectors},
 volume = {32},
 year = {1976}
}

@article{kruskal1964multidimensional,
 author = {J. B. Kruskal},
 journal = {Psychometrika},
 pages = {1--27},
 publisher = {Springer-Verlag},
 title = {Multidimensional scaling by optimizing goodness of fit to a nonmetric hypothesis},
 volume = {29},
 year = {1964}
}

@article{robinson2017sheaves,
 author = {M. Robinson},
 journal = {Information Fusion},
 pages = {208--224},
 publisher = {Elsevier},
 title = {Sheaves are the canonical data structure for sensor integration},
 volume = {36},
 year = {2017}
}

@article{studer2018channel,
 author = {C. Studer and S. Medjkouh and E. Gonulta{\c{s}} and T. Goldstein and O. Tirkkonen},
 journal = {IEEE Access},
 pages = {47682--47698},
 publisher = {IEEE},
 title = {Channel charting: Locating users within the radio environment using channel state information},
 volume = {6},
 year = {2018}
}

@inproceedings{venna2001neighborhood,
 author = {J. Venna and S. Kaski},
 booktitle = {International conference on artificial neural networks},
 organization = {Springer},
 pages = {485--491},
 title = {Neighborhood preservation in nonlinear projection methods: An experimental study},
 year = {2001}
}

@inproceedings{vindas2024multi,
 author = {Y. Vindas and M. Guillaud},
 booktitle = {IEEE 25th International Workshop on Signal Processing Advances in Wireless Communications},
 pages = {826--830},
 title = {Multi-site wireless channel charting through latent space alignment},
 year = {2024}
}

@inproceedings{wang2008manifold,
 author = {C. Wang and S. Mahadevan},
 booktitle = {Proceedings of the 25th international conference on Machine learning},
 pages = {1120--1127},
 title = {Manifold alignment using procrustes analysis},
 year = {2008}
}
